\documentclass{PoS}
\usepackage{amsmath}
\usepackage{graphicx}
\usepackage{epstopdf}
\def\procspie{\ref@jnl{Proc.~SPIE}}   

\usepackage{soul}
\usepackage{float}
\usepackage{fixltx2e}

\usepackage{eurosym}
\usepackage{caption}
\usepackage{subcaption}

\title{Light-Trap: A SiPM Upgrade for Very High Energy Astronomy and Beyond}

\ShortTitle{Light-Trap: A SiPM Upgrade for Very High Energy Astronomy and Beyond}

\author{\speaker{D. Guberman}$^{1}$, J. Cortina$^{1}$, J. E. Ward$^{1}$, A. Hahn$^{2}$, D. Mazin$^{2}$, J. Boix$^{1}$, A. Dettlaff$^{2}$, D. Fink$^{2}$, J. Gaweda$^{1}$, W. Haberer$^{2}$, J. Illa$^{1}$, J. Mundet$^{1}$, Y. Vera$^{1}$ and H. Wetteskind$^{2}$ \\
       $^{1}$Institut de Fisica d'Altes Energies (IFAE), The Barcelona Institute of Science and Technology, Campus UAB, 08193 Bellaterra (Barcelona), Spain\\
       $^{2}$ Max-Planck-Institut f\"ur Physik, D-80805 M\"unchen, Germany \\
        E-mail: \email{dguberman@ifae.es}}

\abstract{With the development of the Imaging Atmospheric Cherenkov Technique (IACT), Gamma-ray astronomy has become one of the most interesting and productive fields of astrophysics. Current IACT telescope arrays (MAGIC, H.E.S.S, VERITAS) use photomultiplier tubes (PMTs) to detect the optical/near-UV Cherenkov radiation emitted due to the interaction of gamma rays with the atmosphere. For the next generation of IACT experiments, the possibility of replacing the PMTs with Silicon photomultipliers (SiPMs) is being studied. Among the main drawbacks of SiPMs are their limited active area (leading to an increase in the cost and complexity of the camera readout) and their sensitivity to unwanted wavelengths. Here we propose a novel method to build a relatively low-cost pixel consisting of a SiPM attached to a PMMA disc doped with a wavelength shifter. This pixel collects light over a much larger area than a single standard SiPM and improves sensitivity to near-UV light while simultaneously rejecting background. We describe the design of a detector that could also have applications in other fields where detection area and cost are crucial. We present results of simulations and laboratory measurements of a pixel prototype and from field tests performed with a 7-pixel cluster installed in a MAGIC telescope camera.}

\FullConference{35th International Cosmic Ray Conference -- ICRC217-\\
		10-20 July, 2017\\
		Bexco, Busan, Korea}

\begin{document}

\section{Introduction}
\label{sec:intro}

After thirty years of constant development Very High Energy (VHE, E $>$ 50 GeV) gamma-ray astronomy has become a mature and productive field in astrophysics. It relies on the Imaging Atmospheric Cherenkov Technique (IACT), which uses one or several telescopes (optical reflectors with fast electronics) to measure the Cherenkov light flashes produced by gamma-ray initiated extensive air showers (EAS) in the upper atmosphere. The detection of those flashes is extremely challenging because of their very short duration ($\sim$3~ns) and their low light intensity. Fast photodetectors with a wide dynamic range and good single photoelectron resolution would be extremely beneficial. Most of the current IACT telescopes are equipped photomultiplier tubes (PMTs). The recent developments in silicon photomultiplier (SiPMs) threatens the hegemony of PMTs in the field. FACT is the first telescope fully equipped with SiPMs and has already been operating for several years\cite{FACT}. The use of SiPMs is also being considered for the next generation CTA observatory and several studies are currently ongoing\cite{SiPM1,SiPM2,SiPM3,SiPM4,LTiWorid}.

\begin{figure}
\centering
\includegraphics[width=0.6\columnwidth]{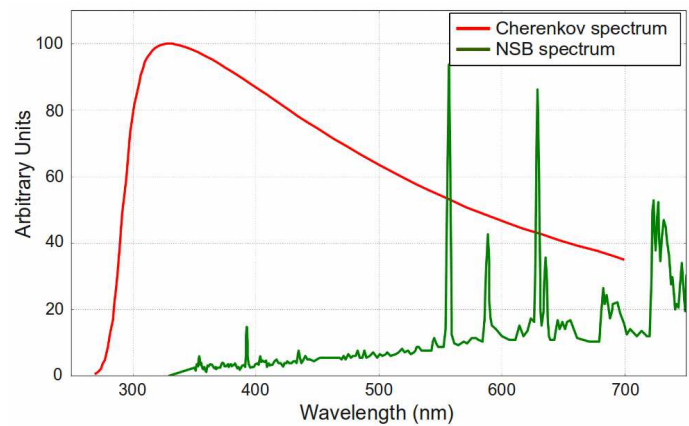}
\caption{The Night Sky Background (NSB) \cite{a} and Cherenkov spectrum at 2200 meter altitude in La Palma, Spain \cite{b}.}\label{fig:NSB}
\end{figure}

Some of the future physics goals in VHE astronomy will require the ability of IACT telescopes to observe several-degree sections of the sky at one time \cite{CTA, machete}. To achieve these goals, more and larger cameras would need to be built. Cost of larger cameras would undoubtedly increase and it may even become prohibitive if utilising PMTs at $>$ \EUR{100} /pixel. At the same time, the cost of SiPMs is decreasing and hence slowly replacing PMTs with SiPMs appears as the most promising approach to solve this dilemma. Besides the cost, SiPMs have several advantages over PMTs: they are robust detectors that can be operated under moonlight without risk of being damaged, can be calibrated easily due to their excellent single photoelectron resolution and are not operated under high voltage. However SiPMs are currently not perfectly suited for VHE astronomy, as individual units are not commercially available in sizes larger than 9$\times$9 mm$^{2}$, they still cost $>$1 \euro{}/mm$^{2}$, reach their PDE at a wavelength longer than the peak emission wavelength of Cherenkov radiation in EAS (see Figure \ref{fig:NSB}), and have a too high PDE at longer wavelengths where there is much less signal from EAS but a large contribution from the Night Sky Background (NSB).

\section{The Light-Trap}

\begin{figure}[t]
  \centering
  \subcaptionbox{Top down view\label{fig:LTTop}}{\includegraphics[width=0.49\columnwidth]{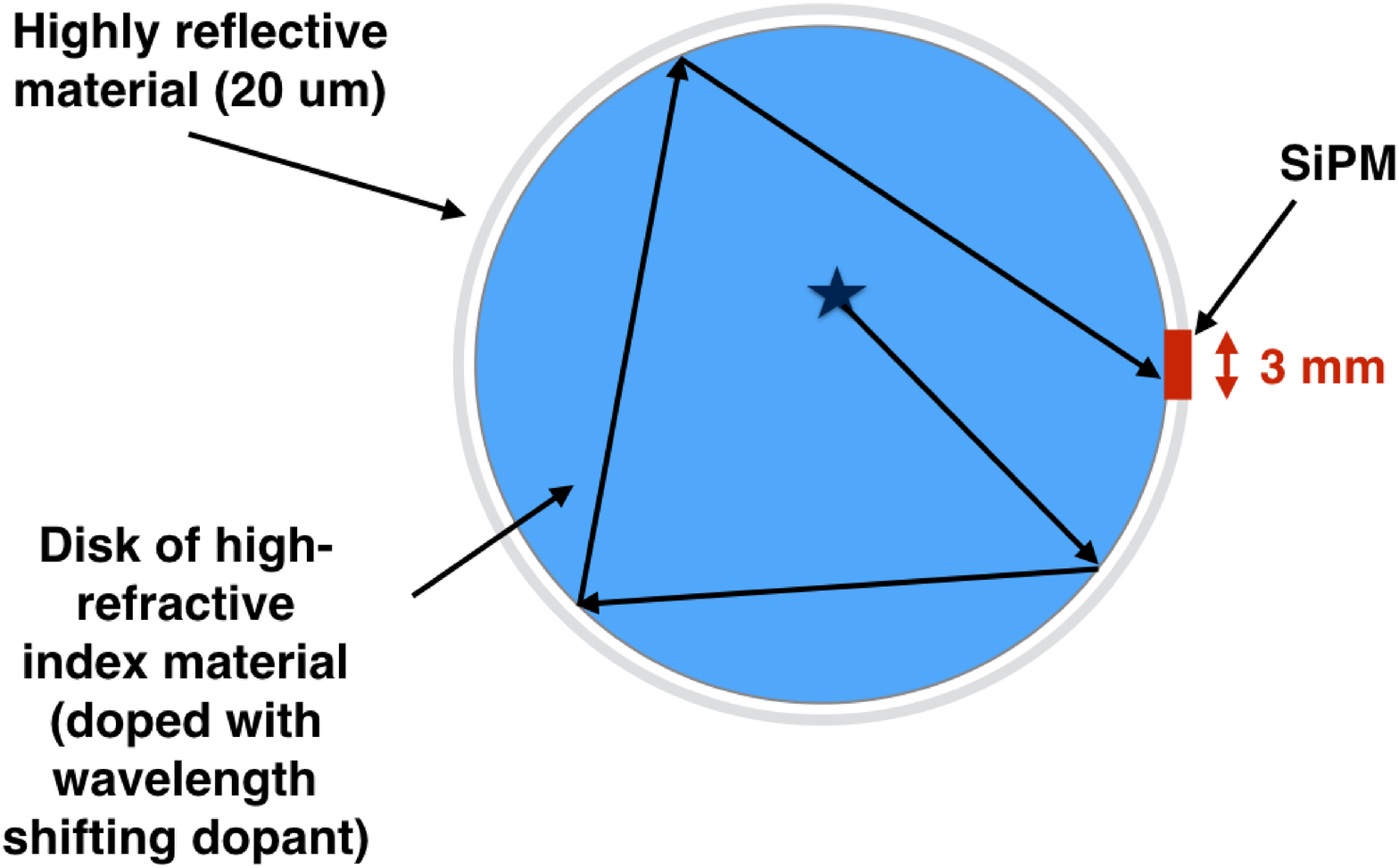}}\hspace{-0.0em}
  \subcaptionbox{Side view\label{fig:LTSide}}{\includegraphics[width=0.49\columnwidth]{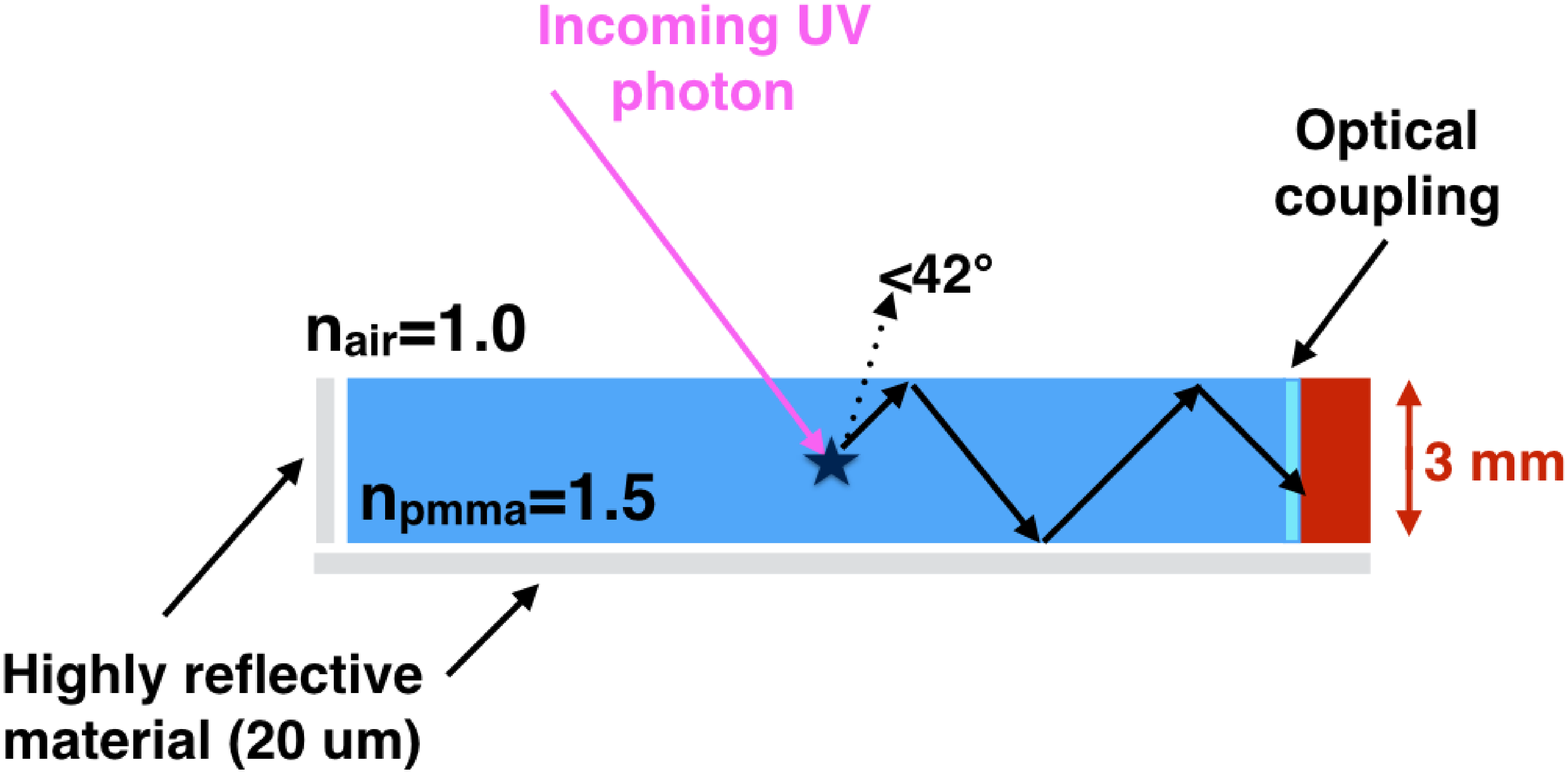}}
  \caption{Conceptual design of the Light-Trap.}
  \label{fig:LTconcept}
\end{figure}

The goal was to build a cheap detector based on SiPM technology with a large detection area, with increased sensitivity for Cherenkov light and that rejects a large fraction of NSB. The \textit{Light-Trap} detector consists on a SiPM coupled to a polymethylmethacrylate (PMMA) disc (see Figure \ref{fig:LTconcept}). This disc contains some wavelength-shifting (WLS) fluors which absorbs photons in the \mbox{$\sim$300-400 nm} wavelength range and re-emits them in the $\sim$400-500 nm range. WLS photons are re-emitted isotropically, so a fraction of them  are trapped in the disk by total internal reflection (TIR) and eventually reach the SiPM. The rest of the re-emitted photons escape. Some of the photons that escape through the side-walls or the bottom can be recovered with the addition of reflective surfaces near the disc.

As a result,

\begin{enumerate}
  \item Light around the peak of the Cherenkov spectrum ($\sim$350 nm) is collected.
  \item Light at longer wavelengths (for which the NSB dominates) is not absorbed by the WLS material and can hardly reach the SiPM.
  \item The absorbed Cherenkov photons are re-emitted at a wavelength where the SiPM PDE is higher
  \item The collection area of the detector can be a factor $\sim$10-50 larger than the sensitive area of the SiPM, i.e. the cost is reduced by the same factor (if the cost of the disk is low) and thus enables us to build pixels far larger than commercially available SiPMs.
\end{enumerate}

\section{Proof-of-Concept Pixel}

\begin{figure}[t]
 \centering
\includegraphics[height=4cm]{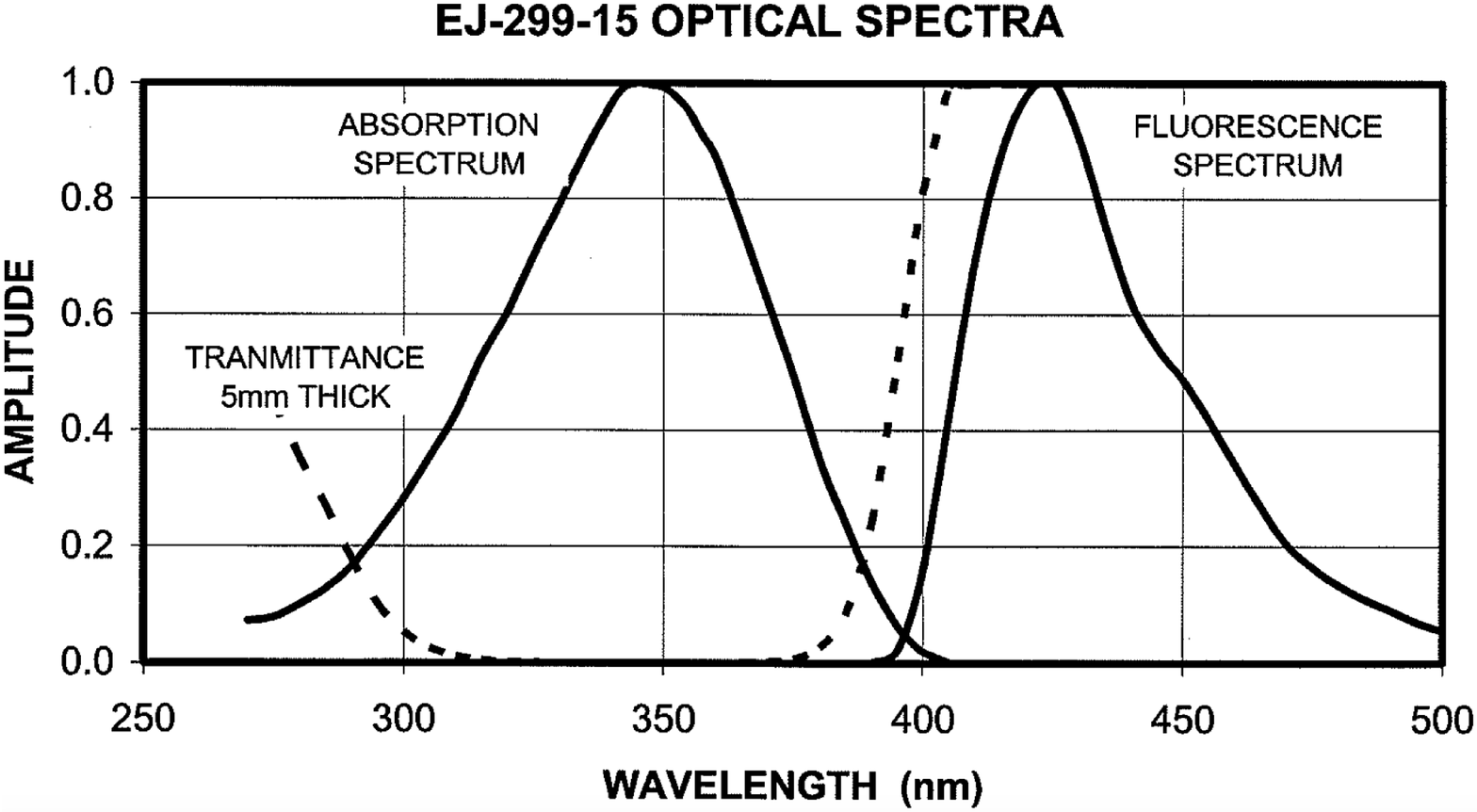}
\includegraphics[height=4cm]{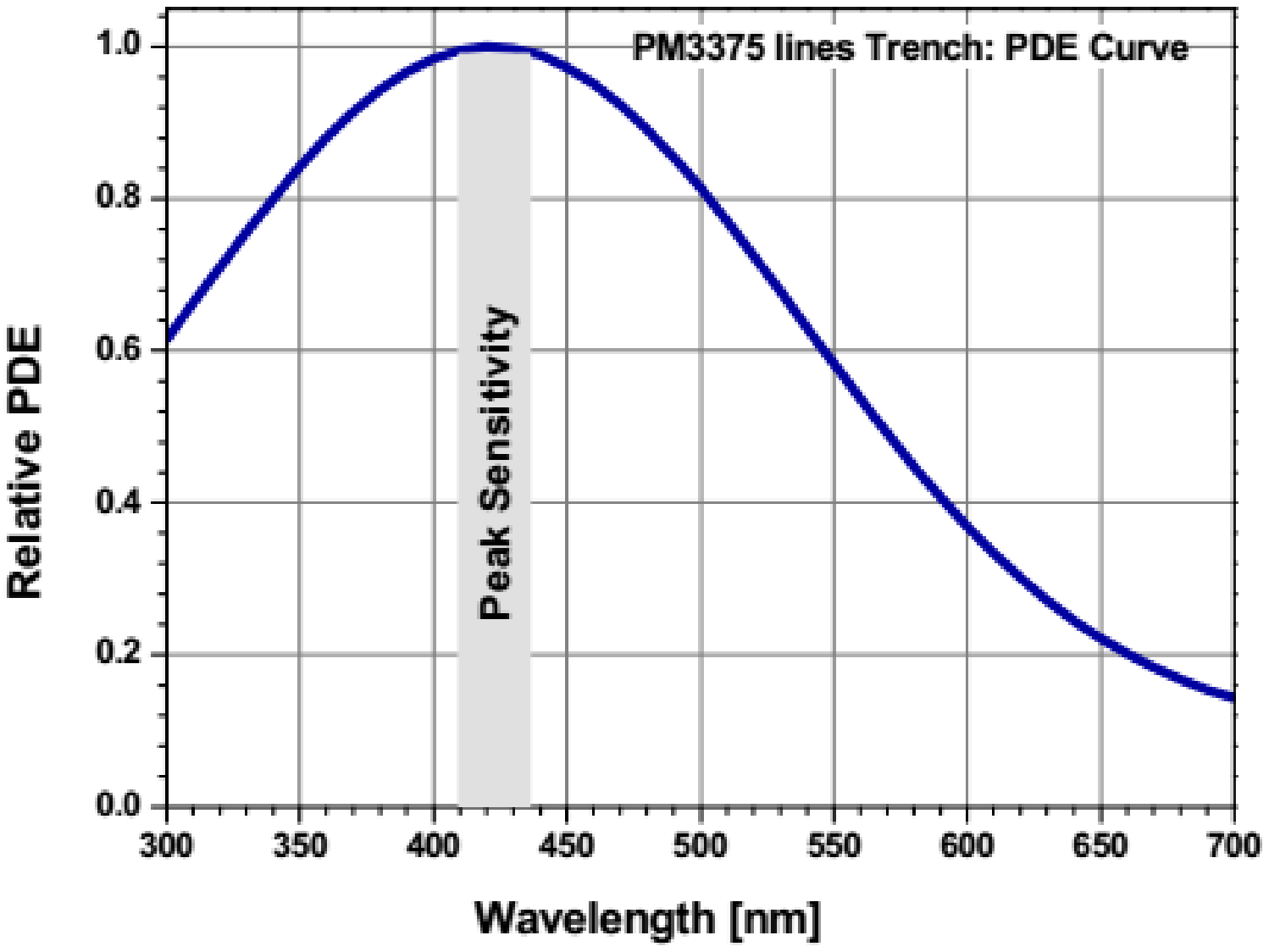}
  \caption{\textbf{Left:} EJ-299-15 absorption and emission spectra, as provided by the company. \textbf{Right}: Relative PDE for the KETEK SiPM used, as provided by the company.}\label{fig:EJ29915}
\end{figure}

To produce a proof-of-concept device, custom-doped wavelength-shifting PMMA (refractive index $n=1.49$) disks were purchased from Eljen Technology (Sweetwater, Texas). We purchased disks of 15~mm diameter with a 3~mm thickness doped with EJ-299-15, which absorbs in the UV band and re-emits in blue-violet (see left panel in Figure \ref{fig:EJ29915}). The dopant levels were customised by Eljen according to our specifications, i.e. to absorb 100\% of incident 340 nm photons within 1.5 mm of the material. The wavelength-shifting fluors added to the disks also have fast responses on the order of $\sim$1 ns, with a quantum yield of $\sim$84\%. 

The SiPM used for this study was the 3~mm~$\times$~3~mm KETEK PM3375 (with a peak sensitivity at 420 nm, see right panel in Figure \ref{fig:EJ29915}), which was coupled to the disk using an optically clear silicone rubber sheet (EJ-560, also purchased from Eljen Technology) that was cut to the SiPM size accordingly. This silicone material is soft and only lightly adhesive, thus allowing the removal and addition of the SiPM. To help improve efficiency in the event that wavelength-shifted photons are not total internal reflected or escape due to optical imperfections, 3M\textsuperscript{\textregistered} ERS reflective foil was cut so as to surround the back and sides of the disk. 

\begin{figure}
   \centering
   \includegraphics[width=0.5\columnwidth]{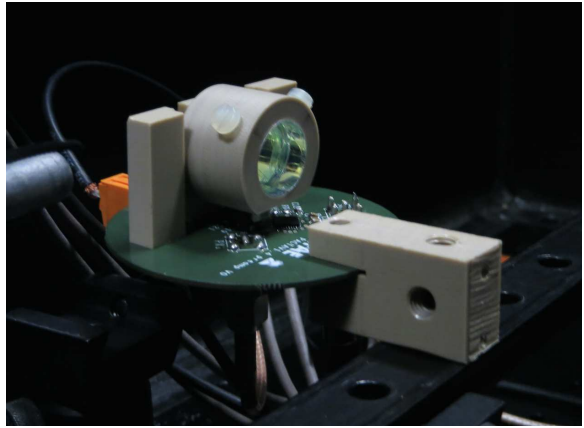} 
   \caption{Image of the Light-Trap holder and electronics board. The SiPM is facing upwards in this picture and the disk surface is pointing to the right.}
   \label{fig:LTHold}
\end{figure}

In order to hold the SiPM, PMMA disk, and reflective foil together it was necessary to construct a cylindrical polyethylene holder (see Figure \ref{fig:LTHold}).

We tested the system by flashing the Light-Trap with four fast-response LEDs of different colours, peaking at 375, 445, 503 and 600 nm. A simple ratio of the Light-Trap output signal to the one obtained with the same SiPM without the disk (\textit{naked-SiPM}) allowed for an estimation of the "boost" factor achieved by the additional use of the scintillating PMMA disk. The ratio of this boost factor with that expected by the simple geometric consideration (i.e. the increase in area between the 9 mm$^{2}$ KETEK SiPM and 176.71 mm$^{2}$ disk, a factor 19.63) gives the ``trapping efficiency'' of the Light-Trap device. 

To help understand the performance and characteristics of the Light-Trap, a simple simulation of the system was created using the Geant4 software simulation package (version 4.10.01-p02\cite{b}). A perfectly optically polished 15~mm disc with the absorption and emission properties of the EJ-299-15 material was simulated. A coupling material (PMMA, with no dopant) was also added to optically couple a sensitive detector to the disk (i.e the SiPM, but the properties of the KETEK device have not been included in these simulations). Finally, two mirrors with 99\% reflectivity were placed at the bottom and sides of the disk, while leaving a 20~$\mu$m air gap at the bottom and 100~$\mu$m at the sides.

\begin{figure}
   \centering
   \includegraphics[width=0.9\columnwidth]{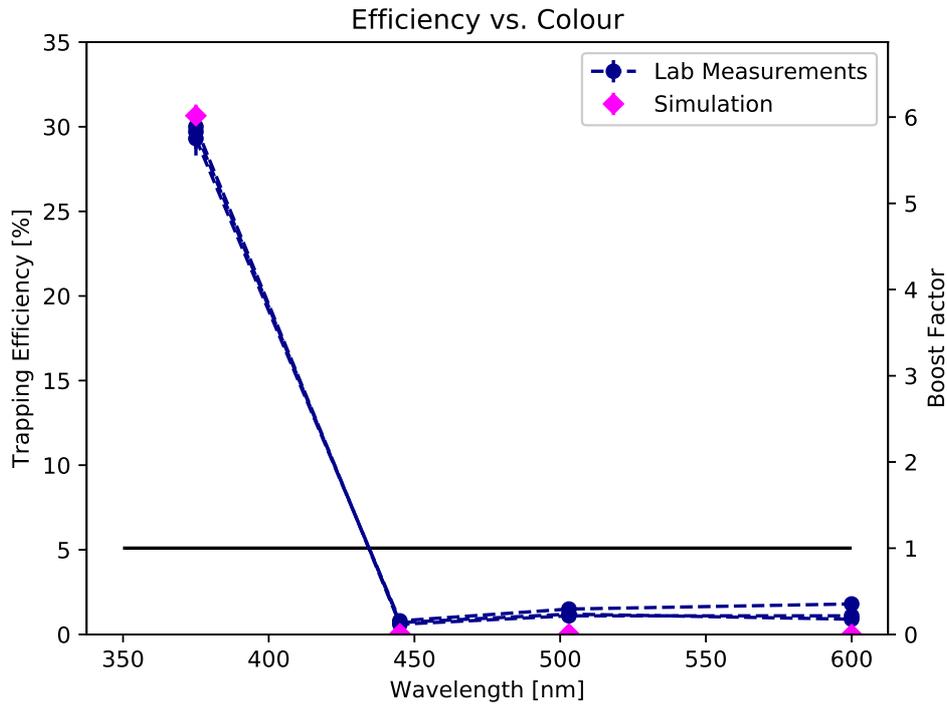} 
   \caption{Trapping efficiency and boost factor of the Light-Trap proof of concept pixel as a function of wavelength.}
   \label{fig:EffBoost}
\end{figure}

The results comparing lab measurements and simulations are shown in Figure \ref{fig:EffBoost}. The detector tested collects the same amount of UV light that what would be achieved using six SiPMs similar to the one used to build thee Light Trap, which is equivalent to a detection efficiency of $\sim$30\%. The detector is almost blind to longer wavelengths. The obtained results agree with the simulations performed with  Geant4.

\section{Field tests at the MAGIC telescopes}

To test the performance of the detector on field  a 7-pixel Light-Trap cluster that was designed, produced and installed in the camera of one of the MAGIC telescopes\cite{upgrade1}. As it had to be adapted to occupy the space of a MAGIC PMT, we purchased from Eljen Technology a set of PMMA discs similar to the ones used for the proof-of-concept pixel, but this time of 20~mm diameter. Surface mount SensL MicroFJ-SMA-30035 SiPMs were used for the cluster. Instead of using 3M\textsuperscript{\textregistered} ERS reflective foil as in the previous section, we built a pixel holder made of highly diffusive material, namely olytetrafluoroethylene.

\begin{figure}[t]
\centering
\includegraphics[height=5cm]{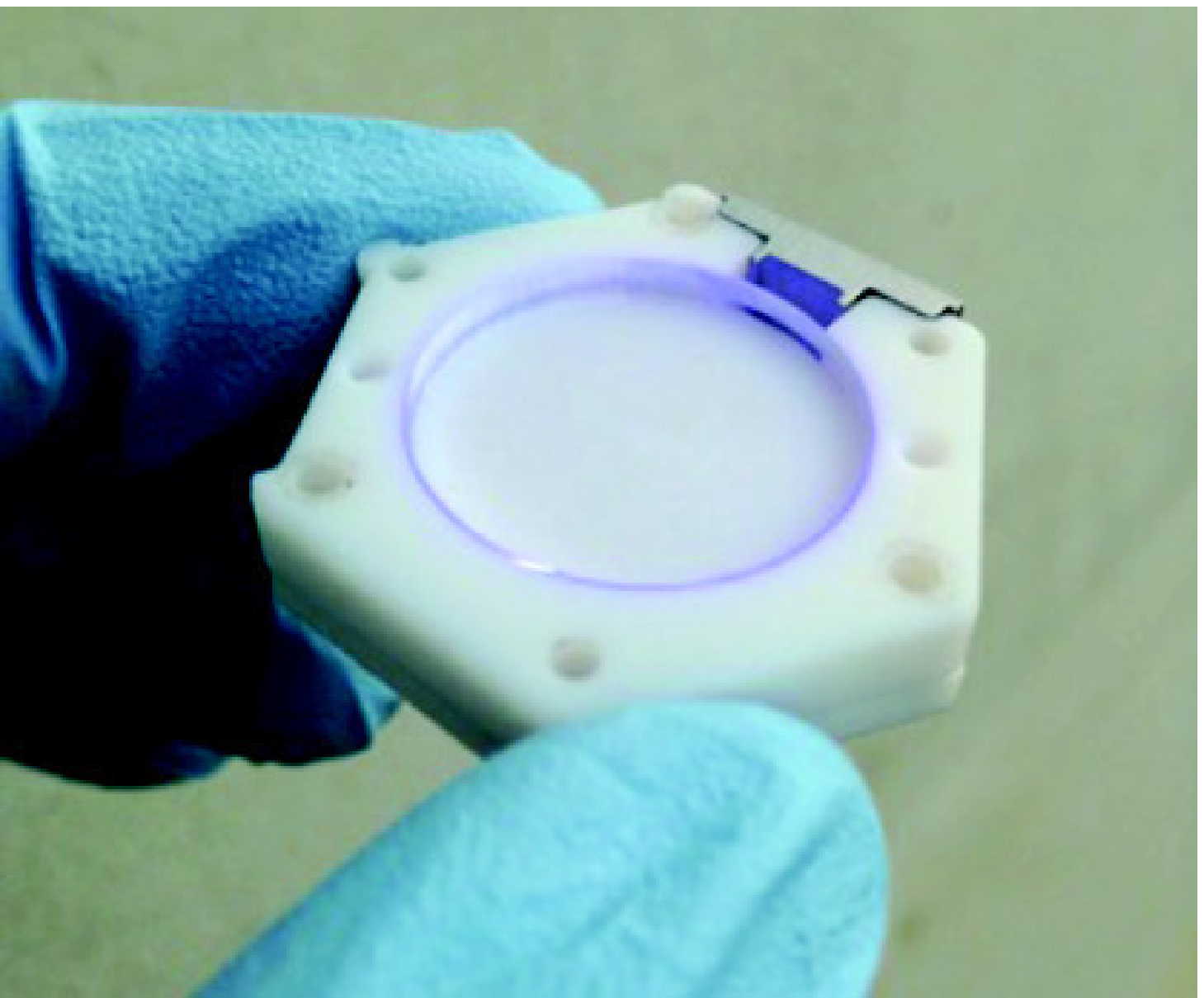}
\includegraphics[height=5cm]{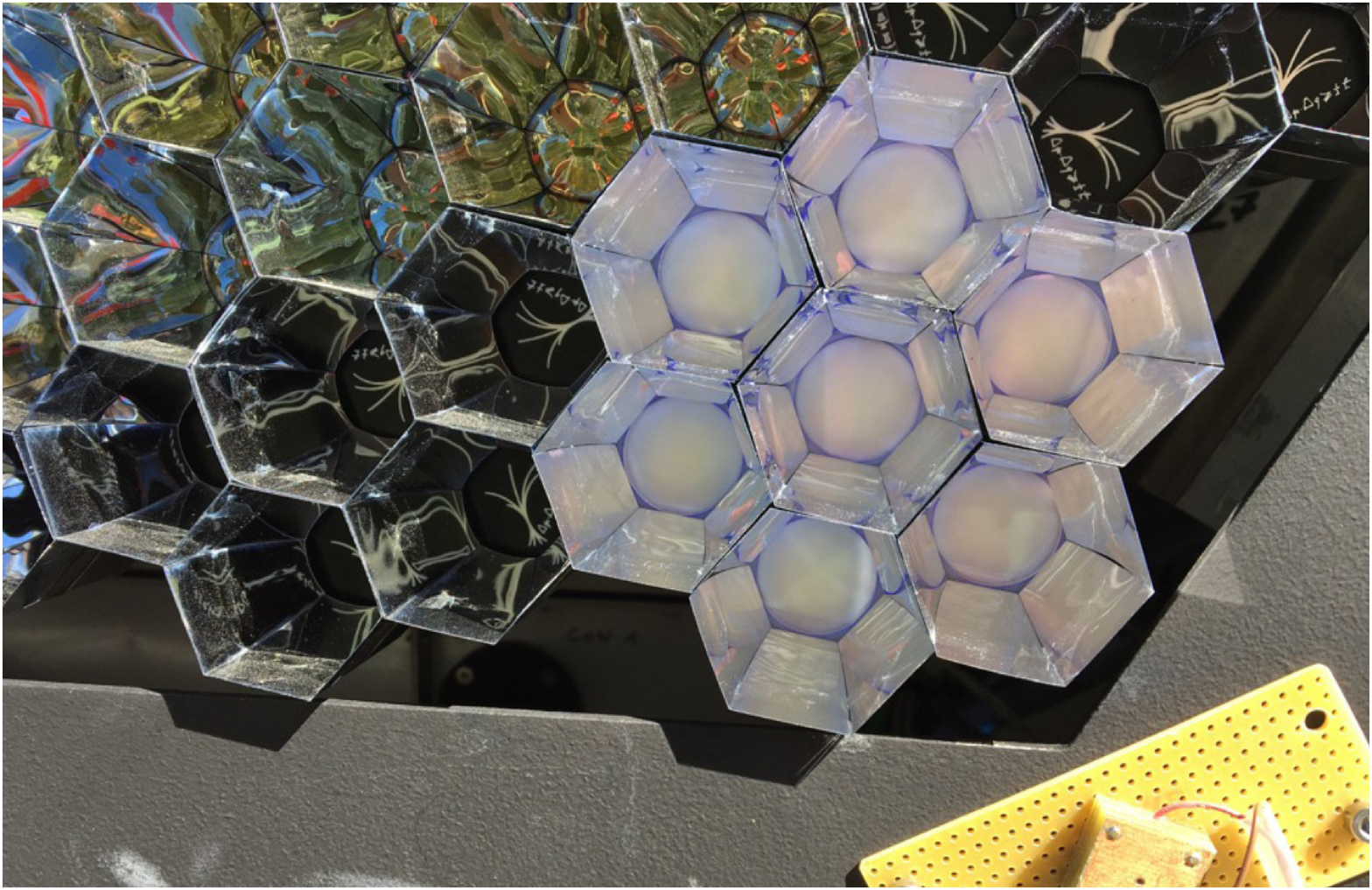}
\caption{\textbf{Left:} One of the pixels of the Light-Trap cluster. \textbf{Right:} The Light-Trap cluster installed in one of the edges of MAGIC 1 camera.}\label{fig:cluster}
\end{figure}

The cluster was installed in the camera of the MAGIC 1 telescope\cite{upgrade1} in May 2017 and it has been operating since then. Tests to evaluate its performance are currently ongoing. A picture of a cluster pixel and one of the cluster installed in the camera are shown in Figure \ref{fig:cluster}.

\section{Conclusions}

A novel method to build relatively low-cost SiPM-based pixels utilising a disk of wavelength-shifting material has been proposed, and a proof-of-concept device has been constructed and tested in the lab. Also a 7-pixel Light-Trap cluster has been recently installed in MAGIC 1 camera and its performance is currently being studied.

Initial measurements have shown that the Light-Trap concept works, providing a modest trapping efficiency of $\sim$30\%, but that there are several steps to be undertaken to greatly enhance the performance of such a device. In addition, the concept itself of building a low-cost large-area detector that may also be tuned to be sensitive in a desired wavelength range might open new opportunities not only in IACT astronomy but also in other fields, both in science (e.g astroparticle physics, astrophysics, high energy physics) and industry (e.g. medical applications).

\section*{Acknowledgements}

This project is supported by a Marie Sklodowska-Curie individual European fellowship. The authors also gratefully acknowledge support from the MAGIC collaboration during the installation and testing of the cluster in MAGIC 1 camera.

\end{document}